\documentclass[]{aastex631}

\shorttitle{Type I X-Ray Burst Models With Rotation}
\shortauthors{Martin \& Jos\'e}
\graphicspath{{./}{figures/}}

\begin{document}

\title{Type I X-Ray Burst Models With Rotation}

\author{David Martin}
\affiliation{Departament de F\'\i sica, Universitat Polit\`ecnica de Catalunya \\
Av. Eduard Maristany 16 \\
Barcelona, E-08019, Spain}

\author[0000-0002-9937-2685]{Jordi Jos\'e}
\affiliation{Departament de F\'\i sica, Universitat Polit\`ecnica de Catalunya \\
Av. Eduard Maristany 16 \\
Barcelona, E-08019, Spain}
\affiliation{Institut d'Estudis Espacials de Catalunya \\
C. Esteve Terradas 1 \\
Castelldefels, E-08860, Spain}

\correspondingauthor{Jordi Jos\'e}
\email{jordi.jose@upc.edu}

\begin{abstract}
Type I X-ray bursts are powered by unstable thermonuclear burning on the surface of accreting neutron stars in close binary systems.
These brief X-ray flashes, with light curves featuring rise times of $1 - 10$ s, durations of $10 - 100$ s, and recurrence periods of hours to days,
represent the most frequent stellar explosions in our Galaxy. With typical energies of $\sim 10^{39} - 10^{40}$ erg, they rank among the most powerful
astrophysical transients after supernovae and classical novae. To date, roughly 120 bursting X-ray binaries have been identified in the Milky Way.

Several studies have been conducted to characterize the dynamics of these events, with emphasis on reproducing the observed recurrence periods and
light curve shapes. In this paper we show, for the first time, that rotation is a key factor shaping the properties of Type I X-ray bursts in rapidly spinning systems.
The inclusion of centrifugal forces, together with a suite of rotationally-induced mixing mechanisms, such as meridional circulation and shear-induced turbulent diffusion, reduce surface gravity, shortening the recurrence times and lowering burst energies. Rotation also modifies the extent of the nuclear activity during these events and affects the morphology of their light curves, which are distinctly broader for rapidly rotating neutron stars.
\end{abstract}


\section{Introduction} \label{sec:intro}

Roughly half of the stars in the Milky Way reside in binary or multiple stellar systems, a fraction of which host a compact object. 
Among these, low-mass X-ray binaries consist of a neutron star and a low-mass companion (typically, a main-sequence or, in some cases, a more 
evolved star) in a close orbit with periods ranging typically from about 0.2 to 15 hours. The secondary star overfills its Roche lobe and mass-transfer 
episodes ensue through the inner Lagrangian point of the system. The hydrogen- and helium-rich material stripped from the secondary carries 
angular momentum and forms an accretion disk orbiting around the neutron star. Viscous and dissipative processes within the disk drive part 
of the material to spiral inward and pile up on the neutron star surface in mildly degenerate conditions. Compressional heating raises the 
temperature of the envelope until a thermonuclear runaway ensues, powering a transient, X-ray-bright event characterized by a rapid rise in 
luminosity up to $\sim 10^4 - 10^5$ L$_\odot$ \citep{SB06, KZ08, JOSE16, GK21}.

With a neutron star hosting the explosion, temperature and density in the accreted envelope reach extreme values, with peaks around 
$T_{peak} \sim 10^9$ K and $\rho_{max} \sim 10^6$ g cm$^{-3}$. Under such conditions, nucleosynthesis proceeds through a complex network 
involving hundreds of isotopes linked by thousands of nuclear interactions, extending all the way up to the SnSbTe-mass region \citep{SCHATZ01a} 
or even beyond (see \citealt{Koi04}, for simulations reaching $^{126}$Xe). Although it is generally accepted that the nuclear activity in XRBs 
is dominated by charged-particle reactions, the extent of the rapid proton-capture (rp) process is still a matter of debate. In the typical 
mixed H/He ignition regime, the dominant nuclear flow arises from the interplay of the rp-process (rapid proton captures and $\beta^+$-decays), 
the triple-$\alpha$ reaction, and the $\alpha$p-process (a suite of ($\alpha$, p) and (p, $\gamma$) reactions that drive the nuclear activity far 
from the valley of stability and eventually reaching the proton drip line beyond mass $A = 38$; \citealt{Woo04,Fis08,JOSE10}). 
The thermonuclear runaways that power XRBs are typically quenched by fuel exhaustion rather than by envelope expansion, owing 
to the extremely high escape velocity from the neutron star surface ($\sim 200000$ km s$^{-1}$). These events occur at nearly constant pressure at 
the ignition layers and synthesize intermediate-mass and heavy species, mostly around masses $A = 60 - 70$.

The potential contribution of XRBs to the Galactic abundances remains also as an open question. The energy required to escape from the surface of a 
neutron star of mass $M_{NS}$ and radius $R_{NS}$ is about $G \, M_{NS}/R_{NS} \sim 200$ MeV nucleon$^{-1}$, whereas only a few MeV nucleon$^{-1}$ are 
released from thermonuclear fusion of solar composition material, making mass ejection extremely unlikely. Nonetheless, radiation-driven winds during 
photospheric radius expansion may expel a small fraction of the outer envelope containing nuclear-processed material \citep{Ebi83,Kat83,Tur86,PP86,JM87}. 
 Recent radiation-driven wind models coupled to hydrodynamic 
XRB simulations have shown that approximately 0.1\% of the envelope mass, enriched in  $^{60}$Ni, $^{64}$Zn, $^{68}$Ge, $^4$He, and $^{58}$Ni,  
is expected to be ejected in a typical XRB \citep{HSJ23}.

XRBs display a remarkable diversity of light curves, ranging from fast-rise, power-law decay profiles with varying recurrence times to more irregular 
patterns, including double or even triple luminosity peaks. The morphology of these light curves is highly sensitive to the adopted initial conditions, 
including the neutron star mass-radius relation and temperature (or luminosity), as well as the metallicity and mass-transfer rate from the secondary 
star\footnote{A self-consistent analysis of the effect of the uncertainties in the neutron-star mass and radius (or equation of state) on XRB light curves has been performed by \citet{Dohi21}. The impact of the neutron-star temperature (or initial luminosity) has been discussed in \citet{Meisel18, Dohi22,Zhen23}. The effects of the mass-transfer rate and metallicity have been studied in numerous works, including \citet{Woo04,JOSE10}, or more recently, \citet{Johnston20}.}. Despite evidence that some neutron stars rotate at very high angular frequencies, the role of rotation in shaping XRB phenomena has received very 
limited attention to date. Previous studies have examined its influence on the propagation of the ignition front \citep{Cave15,Harpo21}, 
 or on the mass-accretion rate threshold separating unstable (bursting) from stable burning regimes \citep{Keek09}. However, 
no comprehensive analysis has yet addressed how rotation affects the global properties of XRBs, from accretion and ignition to explosion, expansion, 
and the quenching of the thermonuclear runaway. This study aims to bridge this gap.

\section{Model and Input Physics}
\label{tmodel}
We assessed the astrophysical impact of rotation through a set of five  
hydrodynamic models of Type I X-ray bursts. The simulations were carried out using 
{\tt SHIVA}, a one-dimensional, Lagrangian, finite-difference, time-implicit hydrodynamic 
code \citep{JOSE98,JOSE16}. {\tt SHIVA} solves the standard set of 
differential equations governing stellar evolution and has been extensively employed 
for more than 30 years in the modeling of classical and recurrent novae, 
Type I X-ray bursts, and sub-Chandrasekhar supernova explosions. The code uses a 
general equation of state that accounts for contributions from the degenerate electron gas, 
the ion plasma, and radiation. Coulomb corrections to the electron pressure are included, 
and both radiative and conductive opacities are considered in the energy transport.
Nuclear energy generation is computed using a reaction network that includes 325 isotopes 
(ranging from $^1$H to $^{107}$Te) linked by 1392 nuclear processes, with updated reaction rates 
from the STARLIB database (\citealt{Sallaska13}; Iliadis, priv. comm.).

\subsection{Implementation of Rotation}
\label{CH4_Rotation_profiles}

We adopted two fundamental assumptions in this work:
\begin{enumerate}
\item The shellular rotation approximation, following the prescription of \citet{MEYNET97} (see Appendix A).
\item Neglect of angular momentum accretion, i.e., the entropy and angular velocity of the accreted material are assumed to be identical to those of the neutron star envelope surface prior to the addition of new material.
\end{enumerate}

The implementation and resolution of the advection-diffusion equation governing angular momentum transport poses multiple numerical challenges, particularly when coupled with the structure and nucleosynthesis equations. For this reason, rotational effects are often restricted to hydrostatic phases, during which the physical variables evolve slowly.
Various approaches have been proposed in the literature to mitigate the numerical difficulties associated with rotation. These include the use of smoothed profiles for the quantities involved in the calculations \citep{Palacios02} and the adoption of fixed rotation profiles once a steady-state rotational regime has been established \citep{DENISSENKOV99,TALON97,MEYNET00}.

For a neutron star envelope, the time required to reach a steady-state rotational regime is extremely short. This relaxation timescale can be roughly estimated a priori as the ratio between the stellar radius and the meridional circulation velocity at the surface, $U(r)$ \citep{ZAHN92}:
\begin{equation}
\label{eq:4Trel}
\tau_{\mathrm{rel}} \approx \frac{R}{U} \approx \tau_{\mathrm{ES}} \propto \tau_{\mathrm{KH}}\,\Omega_\mathrm{S}^{-2} \left( \frac{M^2}{R^3} \right),
\end{equation}
where $\Omega_\mathrm{S}$ is the surface angular velocity, $\tau_{\mathrm{ES}} = \frac{GM^2}{LR} \left( \frac{GM}{\Omega^2 R^3} \right)$ is the Eddington-Sweet timescale, 
and $\tau_{\mathrm{KH}} = \frac{GM^2}{LR}$ is the Kelvin-Helmholtz timescale.

In our models, the resulting relaxation time is approximately $\tau_{\mathrm{rel}} \sim 10^{-19}\ \mathrm{s}$.
This reveals the extremely rapid convergence times arising from the high rotation rates, thin envelope shells, and the extreme temperature and density conditions at the surface of a neutron star. 
Consequently, constant rotation profiles were adopted in our simulations once a steady-state rotational regime was reached.

\subsection{Initial Rotational Profiles}

The rotational conditions adjust during the early stages of the simulations until a steady-state rotational regime is established throughout the envelope.
During this transient phase, the angular velocity profile $\Omega(r)$ changes only slightly from its initial configuration, decreasing at the stellar surface by approximately 3 parts in 10 million across all models considered 
(see Fig.~1).
This results in an extremely mild redistribution of angular momentum within the envelope, driven by meridional circulation and shear-induced turbulent diffusion.
Therefore, it can be concluded that the envelope of a neutron star rotates with an almost uniform angular velocity, following a local solid-body rotation law.
This finding is consistent with \citet{kippenhahn1967} and \citet{MEYNET97}, who argued that the envelopes of massive, hot stars, containing only a small fraction of the total stellar mass, rotate with a uniform angular velocity equal to that of the first envelope shell.

\begin{figure}[h]
\label{Conv1}
\centering
\includegraphics[width=0.8\textwidth]{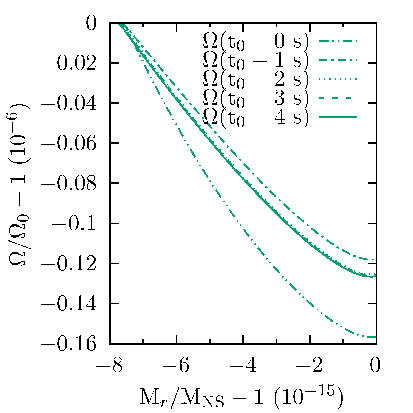}
\caption{Convergence of the initial angular velocity profile, $\Omega(r)$, toward the steady-state regime for Model~2, with an initial value 
of $\Omega_0 = 0.2 \, \Omega_{\mathrm{crit}} = 1.510~\mathrm{rad \, s^{-1}}$, across the envelope. The surface of the accreted envelope corresponds to $M_r/M_{\mathrm{NS}} - 1 = 0$. The asymptotic steady-state solution is shown by the solid line.}
\end{figure}

\begin{figure}[h]
\label{Conv2}
\centering
\includegraphics[width=0.8\textwidth]{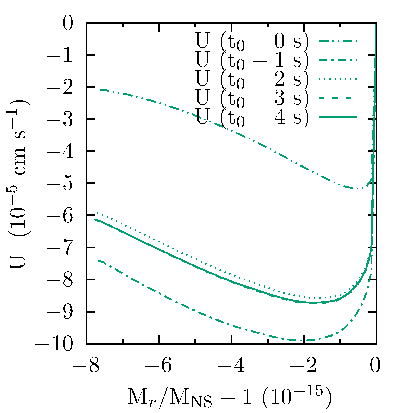}
\caption{Same as Fig. 1, 
for the convergence of the initial meridional circulation velocity profile, $U(r)$.}
\end{figure}

Since meridional circulation and shear transport depend on the degree of differential rotation, the small variations in angular velocity translate into a modest vertical component of the meridional circulation velocity (Fig.~2). 
This component reaches $U(r) \approx -4\times10^{-5}\ \mathrm{cm \, s^{-1}}$ for Model~2 and $\approx -1.2\times10^{-4}\ \mathrm{cm \, s^{-1}}$ for Model~5.  Such circulation velocity is negative throughout the envelope, indicating a flow directed downward along the polar axis and upward along the equatorial plane. Consequently, angular momentum is transported outward, with $U(r)$ decreasing radially and steepening in the outermost layers of the envelope.
The Gratton - \"Opik term, $-\frac{\Omega^2}{2\pi G \rho}$, included in the expression for $U(r)$ (i.e., Eq.~A31 in Appendix A), provides a physical explanation for this behavior: as the density drops significantly in the outer regions of the envelope, the contribution of this term becomes dominant, leading to an inverse circulation pattern. This results in a large circulation cell descending at the poles and rising outward along the equatorial plane.

Figs.~3 and 4 
 show the equilibrium solutions obtained after convergence. The shapes of the resulting profiles are consistent with those reported by 
\citet{MEYNET00} for 20~$\mathrm{M}_\odot$ stellar models with solar metallicity. However, their circulation velocities were on the order of $U(r) \approx 10^{-2}\ \mathrm{cm \, s^{-1}}$, whereas the values obtained in this work for neutron star envelopes are about $U(r) \approx 10^{-5}\ \mathrm{cm \, s^{-1}}$.  \citet{Marques13} reported similarly small values in the central regions of 5~$\mathrm{M}_\odot$ stars, and \citet{TALON97} also found very low circulation velocities ($U(r) \approx 10^{-5}\ \mathrm{cm \, s^{-1}}$) for a 9~$\mathrm{M}_\odot$ stellar model.
We therefore conclude that the circulation velocities derived in this study are consistent with values reported in the literature, despite the significantly higher angular velocities associated with the rapidly rotating neutron star models considered here.

\begin{figure}[h]
\label{Conv3}
\centering
\includegraphics[width=0.8\textwidth]{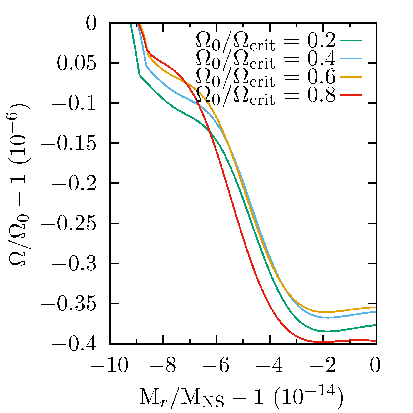}
\caption{Steady-state angular velocity profiles for all models computed in this work. Values are shown relative to the initial uniform rotation profile.}
\end{figure}

\begin{figure}[h]
\label{Conv4}
\centering
\includegraphics[width=0.8\textwidth]{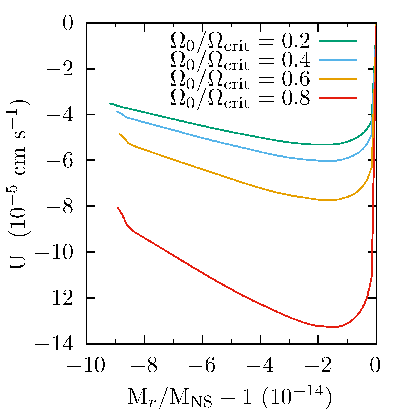}
\caption{Same as Fig. 3, 
for the steady-state meridional circulation velocity profiles.}
\end{figure}

The departure from sphericity in the stellar structure equations for a rotating neutron star envelope is governed by the factors $f_P$ and $f_T$ (see Appendix A). The simulations presented in this work indicate that deviations from spherical symmetry (i.e., when $f_P$ departs significantly from unity) are negligible for $\Omega_0 / \Omega_{\mathrm{crit}} < 0.4$.

Rotationally-induced mixing driven by meridional circulation and shear-induced turbulent diffusion has also been included. Several prescriptions for the diffusion coefficients associated with turbulence, both in the vertical and horizontal directions, and for the mixing processes driven by rotational instabilities have been proposed in the literature, with varying degrees of complexity \citep{ZAHN92,TALON97,PALACIOS03,TALON08}. The topic remains under debate, as no characterization of these diffusion coefficients can be made from first principles. In this work, we adopt the prescription of \citet{MAEDER03} and \citet{Mathis04} for the coefficient of horizontal turbulent diffusion, $D_\mathrm{h}$ (see Appendix A).
The large disparity in magnitude between the horizontal and vertical turbulent diffusivity coefficients, $D_\mathrm{h}$ and $D_\mathrm{s}$ (the latter dominated by shear instabilities), reinforces Zahn's hypothesis of shellular rotation. Indeed, the dominance of horizontal turbulence suppresses the influence 
of the so-called $\mu$-currents and reduces the driving of meridional circulation.

\section{Results}

The most relevant results of a series of hydrodynamic simulations of type I X-ray bursts, aimed at testing the effects of rotation, are summarized in this Section. To this end, five models 
of 1.4 M$_\odot$ neutron stars (with a radius of R$_{\rm NS}$ = 13.1 km and an initial luminosity of L$_{\rm NS}$ = 4.14 L$_\odot$), accreting solar composition material (Z = 0.02) from a companion star at a rate of $\dot M = 1.75 \times 10^{-9}$ M$_\odot$ yr$^{-1}$ (0.08 $\dot M_{\rm Edd}$) have been considered. The models differ only on the adopted rotational velocity, ranging from 0 (non-rotating Model 1) to 
$\Omega_0 = 6040$ rad s$^{-1}$ (0.8 $\Omega_{\rm crit}$; Model 5; see Table 1 in Appendix B, for further details).
All metals have been initially assumed to be in the form of $^{14}$N, following the rapid rearrangement of CNO isotopes that naturally occurs early in the burst \citep{Woo04}. In fact, the non-rotating Model 1 is similar to Model ZM, computed by \citet{Woo04} with the one-dimensional, hydrodynamic code {\tt KEPLER}, and to Model 1, computed by \citet{JOSE10} with the code {\tt SHIVA}.

\subsection{Non-rotating Model 1}
\subsubsection{First Burst}
During the accretion stage the envelope is progressively compressessed and heated by the accumulation of solar composition material onto the neutron star. 
Nuclear reactions set in near the envelope base, and convection establishes erratically, for the first time, when T$_{\mathrm{base}}$ reaches $3.9 \times 10^{8}$ K, well above the core-envelope interface and progressively extending throughout the whole envelope.
When a critical amount of matter is piled up on top the neutron star, that depends on the mass and temperature -or luminosity- of the neutron star, as well as on the metallicity and mass-accretion rate from the companion star, a thermonuclear runaway (TNR) ensues. 

Maximum density ($\rho_{\mathrm{max,base}} \sim 3.4 \times 10^5$ g cm$^{-3}$) and pressure (P$_{\mathrm{max,base}}=1.2 \times 10^{22}$ dyn cm$^{-2}$) are reached at the base of the envelope, about 5.8 hr (21006 s) since the initiation of accretion, 
when T$_{\mathrm{base}} \sim 2.7 \times 10^8$ K, marking the onset of the expansion stage (envelope size $\Delta z \sim 12.8$ m). The nuclear activity is fully dominated by the CNO-cycle, but leakage from this cycle (mainly through $^{15}$O($\alpha$, $\gamma$)) is progressively increasing the 
metallicity at the innermost envelope shells (from an initial value of 0.02 to 0.15, at this stage). The envelope continues expanding smoothly, nearly at constant pressure. 

At t = 21146 s (5.9 hr), when T$_{\mathrm{base}}$ achieves $1 \times 10^9$ K, the energy generation rate by nuclear reactions reaches its maximum value, $\epsilon_{nuc,max} \sim 4.1 \times 10^{17}$ erg g$^{-1}$ s$^{-1}$. Two seconds later, the envelope attains maximum expansion, with a size $\Delta z_{max} \sim 45$ m. And 4 s later (t = 21150 s), the temperature at the base reaches a maximum value of $T_{peak} \sim 1.07 \times 10^9$ K. Almost simultaneously, the neutron star attains maximum luminosity, L$_{max}$ = $4.0 \times 10^{38}$ erg s$^{-1}$ ($1.0 \times 10^5$ L$_\odot$).

When t = 22910 s (6.4 hr), the overall luminosity of the star has decreased to a minimum value, L$_{NS}$ = $1.0 \times 10^{35}$ erg s$^{-1}$, which somewhat marks the end of the first burst in the simulations. The rate of nuclear energy generation, powered now mostly by $\beta$-decays, has dropped to  $\epsilon_{nuc} \sim 9.3 \times 10^{12}$ erg g$^{-1}$ s$^{-1}$, while the size of the envelope has reduced to $\Delta z \sim 9$ m. At the end of this first burst, the mean, mass-averaged chemical composition of the envelope is mainly dominated by $^{60}$Ni (0.33), $^4$He (0.29), $^1$H (0.17), $^{64}$Zn (0.03), $^{56}$Ni (0.02), $^{52}$Fe (0.02), and $^{12}$C (0.02), with a nuclear endpoint (i.e., the heaviest isotope with a mass-fraction $X > 10^{-9}$) located at $^{89}$Nb. At this stage, the  metallicity of the envelope has reached a mean value of Z $\sim$ 0.54.

\subsubsection{Bursts Second to Fifth}
As discussed in \citet{Taa80} and \citet{Woo04}, {\it compositional inertia}, or the fact that accretion occurs onto the ashes of previous bursts (characterized by the  presence of unburned H and $^4$He), deeply affects the properties of the subsequent bursting sequences. The depletion of H at the innermost envelope layers shifts the location of the ignition region, which progressively moves away from the core-envelope interface in subsequent explosions.
From a nucleosynthesis viewpoint, the nuclear activity extends toward heavier species, attaining endpoints around $^{98}$Ru ($2^{nd}$ burst), $^{101}$Pd ($3^{rd}$ burst), $^{102}$Pd ($4^{th}$ burst), and $^{103}$Ag ($5^{th}$ burst). This, in turn, modifies the overall mass-averaged metallicity of the envelope at the end of each burst [Z = 0.77 ($2^{nd}$ burst), 0.85 ($3^{rd}$ burst), 0.89 ($4^{th}$ burst), and 0.91 ($5^{th}$ burst)]. 
The bursts tend to converge to a constant recurrence period of about $\tau_{rec} \sim$ 5.1 hr, with a ratio between persistent and burst luminosities around $\alpha \sim 31$. The light curves of the five bursts computed for this model are similar, characterized by a fast rise and an exponential-like decay. 

\subsection{XRB Models with Rotation}
\begin{figure}[h]
\centering
\includegraphics[width=0.9\textwidth]{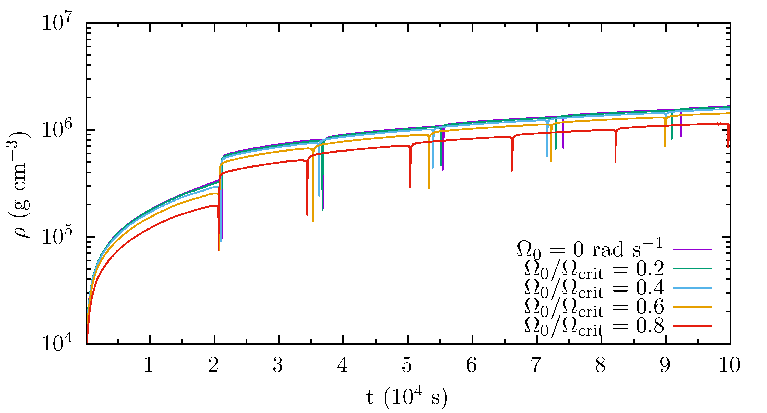}
\caption{Time evolution of the density at the envelope base for all models computed in this work, along the first five bursting episodes. The
origin of the time coordinate is arbitrarily chosen as the time for which $T_{\mathrm{base}} \sim 2.7 \times 10^7$ K.}\label{fig:rho_5b}
\end{figure}

The simulations reported in this work confirm the pressure-lifting effect caused by rotation \citep{KIPP70,Dominguez96}, in the framework of rapidly spinning neutron stars, showing that the  
maximum pressure and density attained at the base of envelope, 
$P_{\mathrm{max,base}}$ and $\rho_{\mathrm{max,base}}$, decrease with increasing spin of the neutron star (see Fig. \ref{fig:rho_5b}). Indeed, at the end of the first burst, $P_{\mathrm{max,base}}$ and $\rho_{\mathrm{max,base}}$ reach values of 
$1.2 \times 10^{22}$ dyn cm$^{-2}$ and $3.4 \times 10^5$ g cm$^{-3}$ (Model 1),
$1.2 \times 10^{22}$ dyn cm$^{-2}$ and $3.3 \times 10^5$ g cm$^{-3}$ (Model 2),
$1.1 \times 10^{22}$ dyn cm$^{-2}$ and $2.9 \times 10^5$ g cm$^{-3}$ (Model 3),
$9.2 \times 10^{21}$ dyn cm$^{-2}$ and $2.5 \times 10^5$ g cm$^{-3}$ (Model 4), 
and 
$6.8 \times 10^{21}$ dyn cm$^{-2}$ and $1.9 \times 10^5$ g.cm$^{-3}$ (Model 5). 
This pressure-lifting effect affects in turn the maximum expansion achieved by the envelope, which experiences a significant growth with the increase of the angular velocity. Indeed, the envelope reaches a maximum size of 44.8 m, 45.9 m, 49.2 m, 56.2 m and 74.4 m, during the first burst, representing an increase of $\approx$ $2\%$, $10\%$, $25\%$, and $66\%$ with respect to the non-rotating model.
Pressure profiles are smooth, with no spikes, confirming that X-ray bursts take place at nearly constant pressure. 

\begin{figure}[h]
\centering
\includegraphics[width=0.9\textwidth]{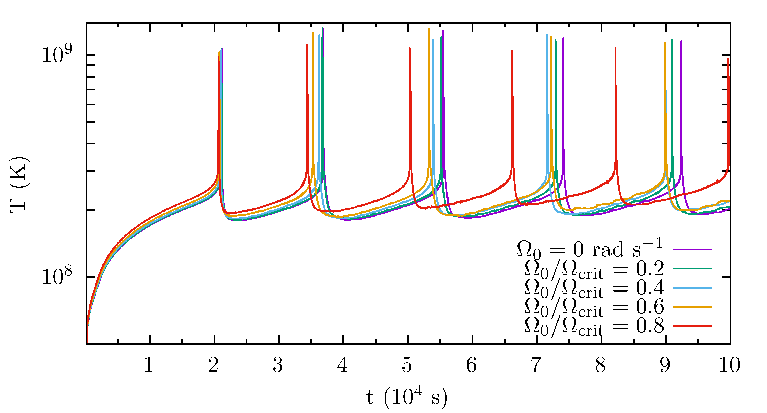}
\caption{Same as Fig. \ref{fig:rho_5b}, for the time evolution of the temperature at the envelope base.}\label{fig:T_5b}
\end{figure}

The recurrence time (or time between two consecutive bursts) is also strongly dependant on the degree of turbulence-induced mixing of chemical species \citep{Fujimoto93}. The reason is simple: if hydrogen and/or helium are transported to greater depths, where temperature and density are higher, thermonuclear reactions will set in sooner. Therefore, rotationally-induced mixing is expected to reduce the recurrence time between successive bursts. Our results, displayed in Fig. \ref{fig:T_5b} in terms of the temperature at the envelope base, confirm this result. 
Bursts for models with higher angular velocities tend to occur earlier than those with less or no rotation. After five bursts, our models show  recurrence times of $\tau_{\mathrm{rec}} = 5.1$ hr (Model 1), 5.0 hr (Model 2), 5.1 hr (Model 3), 4.9 hr (Model 4), and 4.4 hr (Model 5).
It is worth noting that the ignition pressure, the effective gravity, and ignition column density decrease when increasing the neutron-star angular velocity. Since the recurrence time between consecutive bursts, $\tau_{rec}$, is proportional to the ignition column density, for a given mass-accretion rate, $\tau_{rec} \sim y_i \dot{m}$ (where $y_i$ is the ignition column density and $\dot{m}$ is the local mass-accretion rate per unit area), a reduction in $y_i$ translates into a decrease in the recurrence times between bursts. This is in agreement with the results reported in this work (see Table 1).

Models with larger angular velocities tend to reach lower peak temperatures. As shown by \citet{Sha81} and  \citet{Fuj82a}, 
the key factor determining the strength of a thermonuclear explosion is the maximum pressure attained at the base of the envelope, 
$P_{\mathrm{max}}$, which measures the total pressure exerted by the layers overlying the ignition shell:
\begin{equation}
P_{\mathrm{max}} = \frac{G , M_{\mathrm{NS}}}{4 \pi R_{\mathrm{NS}}^4} M_{\mathrm{acc}}
\label{eq:Ptmax}
\end{equation}
Here, $M_{\mathrm{NS}}$ and $R_{\mathrm{NS}}$ are the mass and radius of the neutron star hosting the explosion, 
and $M_{\mathrm{acc}}$ is the mass of the accreted envelope. Equation~\ref{eq:Ptmax} shows that, for a given neutron star, 
$P_{\mathrm{max}}$ depends solely on the accreted mass, which, for a constant mass-accretion rate, is determined by the duration of 
the accretion phase. Models with higher rotational velocities exhibit shorter accretion phases and, consequently, smaller accreted masses. 
This results in less energetic outbursts, characterized by lower peak temperatures.

\begin{figure}[h]
\centering
\includegraphics[width=0.8\textwidth]{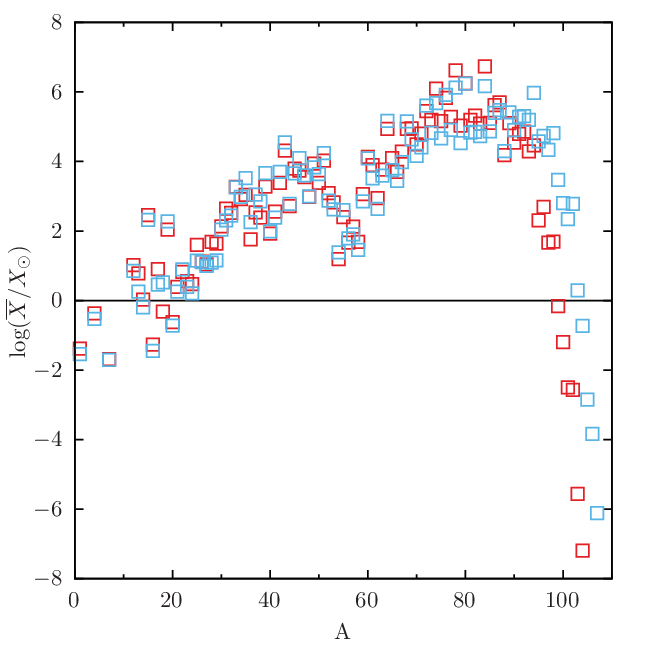}
\caption{Mean overproduction factors of stable isotopes relative to solar abundances (for $f > 10^{-8}$) in Models 1 (blue) and 5 (red), at the end of the fifth 
burst.}\label{fig:of_5b}
\end{figure}

Regarding the associated nucleosynthesis, the $^1$H mass fraction in the innermost shell of the envelope at the onset of the first burst 
is higher in models with faster rotation. This behavior arises from the combined effect of deeper fuel mixing (i.e., rotationally-induced mixing) 
and shorter recurrence times. The resulting reduction in the $^1$H consumption rate in rapidly rotating models leads, in turn, to a smaller 
production of $^4$He.
For Model 1, the mean, mass-averaged chemical composition of the envelope at the end of the fifth burst is dominated by intermediate-mass elements, 
including $^{60}$Ni (0.26), $^{64}$Zn (0.16), $^{32}$S (0.10), and $^{4}$He (0.08). The nucleosynthesis endpoint is located around $^{103}$Ag.
In Models 2 through 5, the most abundant isotopes at the end of the fifth burst are $^{60}$Ni (0.27) and $^{64}$Zn (0.15) for Model 2, 
$^{60}$Ni (0.27) and $^{64}$Zn (0.14) for Model 3, $^{60}$Ni (0.21) and $^{64}$Zn (0.17) for Model 4, and $^{60}$Ni (0.29) and $^{4}$He (0.12) 
for Model 5. The corresponding nucleosynthesis endpoints are $^{102}$Pd for Models 2 to 4, and $^{98}$Ru for Model 5--five mass units lighter than the 
value obtained for the non-rotating Model 1. 
Mean overproduction factors of stable isotopes relative to solar abundances at the end of the fifth are shown, for Models 1 and 5,
in Fig. \ref{fig:of_5b}.

\begin{figure}[h]
\centering
\includegraphics[width=0.8\textwidth]{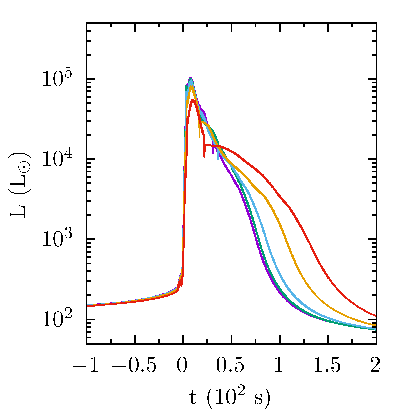}
\caption{Light curves of the first burst, for all models computed in this work. The individual light curves have been
horizontally shifted to align peak values, for a better comparison.}\label{fig:L_5b}
\end{figure}

Although the decline phases of type-I X-ray burst light curves were originally described by a simple exponential function \citep{GRIND76} or 
a power law \citep{Zand14}, more recent studies have revealed a wide diversity of profiles \citep{Kuuttila17}. 
The inclusion of rotation introduces an additional factor influencing the morphology of X-ray burst light curves, as illustrated 
in Fig.~\ref{fig:L_5b}: models with higher rotation rates exhibit not only longer decay times but also markedly altered 
decline shapes. In particular, a noticeable "bump" appears just after the luminosity peak in models with the highest angular velocities. The origin of this bump is not understood: while it does not seem to be related to a drastic opacity change, because of the (time) proximity to the luminosity peak, it may be driven by a nuclear waiting-point impedance in the nuclear reaction flow, causing a deviation from the exponential-like decay pattern (see \citealt{Fis04}, for X-ray burst models with bumps, observable as double-peaked light curves with peaks separated by $\sim$ 5 s). 
Moreover, the larger photospheric expansion of the neutron star envelope driven by a deeper-seated ignition can contribute to the broadening of the bump in the fastest rotating models. Overall, the light curves of rapidly rotating models are broader and deviate significantly 
from a purely exponential decay. Relative to the non-rotating Model 1, burst durations (defined as the time during which $L_{\mathrm{NS}}$ remains 
above 1\% of the peak luminosity) increase by approximately 4\%, 16\%, 46\%, and 86\% for the first burst, for models with $\Omega_0 / \Omega_{\mathrm{crit}} =$ 0.2, 0.4, 0.6, 
and 0.8, respectively; the light curves of the five consecutive bursts computed for Model 5 ($\Omega_0 / \Omega_{\mathrm{crit}} =$ 0.8) are,
respectively about 86\%, 117\%, 125\%, 117\%, and  52\% broader than those corresponding to the non-rotating Model 1. The fact that successive bursts are systematically broader in rapidly rotating neutron-star models proves that this is a true effect induced by rotation rather than a numerical artifact potentially driven by the initial conditions adopted in the simulations.

\section{Discussion}
 This work demonstrates that rotation is a key factor in determining the properties of Type I X-ray bursts in rapidly spinning neutron stars. In the models reported in this work, for neutron stars of 1.4 M$_\odot$, accreting solar composition material from a companion star at a rate of 
$\dot M = 1.75 \times 10^{-9}$ M$_\odot$ yr$^{-1}$,
the centrifugal (lifting) effect, together with the suite of rotationally-induced mixing mechanisms implemented, reduce the maximum density and pressure reached at the base of the accreted envelope, which in turn shorten the recurrence time between bursts.
This results in less energetic bursts characterized by lower peak temperatures.
Models with rapid rotation also exhibit significant envelope expansion, by up to 66\% compared with non-rotating models. 
Rotation further influences the extent 
of the nuclear activity during Type I X-ray bursts, limiting the nucleosynthesis endpoint: differences of up to five mass units are found between the fastest-rotating and non-rotating models.
Perhaps the most striking effect of rotation is seen in the morphology of the light curves, which are distinctly broader for rapidly rotating models, with burst durations 
up to approximately 125\% longer than those of non-rotating counterparts.

It is finally worth noting that Newtonian gravity has been assumed in the simulations reported in this work. However, since the envelope layers considered are very thin,  
general relativity corrections can be introduced in a straightforward form (see \citealt{Ayasli82,Lew93,TAAM93,Cum02,Woo04}, for details). To this end, the star's surface gravity can be rewritten as $g = GM_*/R^2_*(1 + z)$, where $M_*$ is the mass, $R_*$ is the stellar radius (defined in such a way that the surface area of the star is $4\pi R^2_*$), and $z$ is the gravitational redshift given by $1 + z = (1 - 2GM_*/R_*c^2 )^{-1/2}$. Our models of $M_* = 1.4 M_{\odot}$ yield $R_* =$ 14.3 km, and a gravitational redshift of $z = 0.19$.

A suitable observer at infinity would report burst durations and recurrence times increased by a factor of $1 + z = 1.19$. The burst luminosity as well as the mass-accretion rate have to take
into account both the gravitational redshift term and the difference in surface area, compared to the Newtonian framework. The energy and rest mass-accretion rate scale as $R^2_*/(
1 + z)$, while the luminosity $\propto R^2_*/(1 + z)^2$. However, when $M_*$ is taken exactly as $M_{\mathrm{NS}}$ (Newtonian framework), the surface area and redshift corrections for energy and mass-accretion rate cancel each other, since $g \propto (1+z)/R^2_*= const$, and therefore, no correction to the observed burst energy or mass-accretion rate is necessary, while the luminosity correction is given by $1/(1 + z) = 0.84$. 

\begin{acknowledgments}
This paper benefited greatly from insightful feedback provided by
A. Chieffi, R. Hirschi, A. Maeder, and G. Meynet on the implementation of rotation in stellar models. 
This work has been partially supported by the Spanish MINECO grant 
PID2023-148661NB-I00, and by the E.U. FEDER funds.

\end{acknowledgments}


\software{SHIVA \citep{JOSE98,JOSE16}.}

\appendix

\section{Structural Evolution of Rotating Stars} 

Rotation plays a crucial role in stellar evolution, influencing stellar shape, lifetimes, surface properties, and chemical abundances 
\citep{Hunter09}. 
When rotation is taken into account, the mechanical equilibrium of a star is inevitably altered because rotating stars deviate from spherical symmetry. 
In particular:

\begin{itemize}
\item centrifugal forces reduce the effective gravity away from the rotation axis.
\item the radiative flux is proportional to the local effective gravity, for uniformly rotating stars (von Zeipel theorem).
\end{itemize}
The most accurate way to model rotation is through a multidimensional approach. However, an approximate description is still feasible within a one-dimensional (1D) framework. Following \citet{MEYNET97} and \citet{KippenhahnThomas}, we model the structural evolution of rotating stars in the non-conservative case (i.e., the {\it shellular rotation} approximation), in which the angular velocity $\Omega$ depends only on depth, $\Omega(r)$. In this approach, gravitational equipotentials are replaced by isobars, defined as surfaces where the total gravitational potential in the Roche approximation, $\Psi_P$, is constant:
\begin{equation}
\label{eqn:potential}
\Psi_P =-\Phi + \frac{1}{2}\Omega^2r^2 \sin^2 \theta = \mathrm{constant}
\end{equation}
where $\Phi = -G m_P/r$ is the gravitational potential, $G$ is the gravitational constant, $m_P$ is the mass enclosed within a sphere of radius $r$, and $\theta$ is the colatitude. Under these assumptions, the equations governing the structure of a rotating star in hydrostatic equilibrium can be expressed as follows (with the subscript $P$ denoting equipotential surfaces):

\begin{equation}
\label{eqn:massRot}
\frac{\partial r_P}{\partial m_P} = \frac{1}{4\pi r_P^2 \overline{\rho}}
\end{equation}

\begin{equation}
\label{eqn:momentumRot}
\frac{\partial P}{\partial m_P} = - \frac{Gm_P}{4\pi r_P^4}f_P
\end{equation}

\begin{equation}
\label{eqn:energyRot}
\frac{\partial L_P}{\partial m_P} = \epsilon_n - \epsilon_{\nu} + \epsilon_g
\end{equation}

\begin{equation}
\label{eqn:gradTmRot}
\frac{\partial T}{\partial m_P} = -\frac{Gm_PT}{4\pi r^4_P P} \nabla_P
\end{equation}

\begin{equation}
\label{eqn:luminosityRot}
L_P = -\frac{4ac}{3} \langle g^{-1}_{\mathrm{eff}} \rangle S^2_P \langle \frac{T^3g_{\mathrm{eff}}}{\kappa} \rangle \frac{\partial T}{\partial m_P}
\end{equation} 
where $P$, $T$ and $\kappa$ are the pressure, temperature and opacity of the stellar plasma, respectively; $L_P$ is the energy per unit time crossing the equipotential surface $\Psi_P$; $\overline{\rho}$ is the volume-averaged density between two isobars; $\epsilon_n$, $\epsilon_{\nu}$, and $\epsilon_g$ are the nuclear, neutrino, and gravitational energies, respectively; $a$ is the radiation constant, $a = 4 \sigma/c = 7.5657 \times 10^{-15}$ 
erg cm$^{-3}$ K$^{-4}$; and $c$ is the speed of light.
The radiative gradient, $\nabla_P$, in the case of non-conservative rotation is given by:
\begin{equation}
\label{eqn:gradP}
\nabla_P = -\frac{3\kappa}{64\pi \sigma G}\frac{P}{T^4}\frac{L_P}{m_P}\frac{f_T}{f_P}.
\end{equation}
The scalars $f_P$ and $f_T$ are correction factors introduced to recover the original form of the stellar structure equations in the absence of rotation, and are defined as:
\begin{equation}
\label{eqn:f_P} 
f_P = \frac{4\pi r^4_P}{Gm_P S_P}\frac{1}{\langle g_{\mathrm{eff}}^{-1} \rangle}
\end{equation} 
\begin{equation} 
f_T = \left( \frac{4\pi r^2_P}{S_P} \right)^2 \frac{1}{\langle g_{\mathrm{eff}} \rangle \langle g_{\mathrm{eff}}^{-1} \rangle},
\end{equation}
where $\langle g_{\mathrm{eff}} \rangle$ and $\langle g_{\mathrm{eff}}^{-1} \rangle$ are the averages performed on an isobaric surface of the effective gravity $g_{\mathrm{eff}}$, and its inverse $g_{\mathrm{eff}}^{-1}$, respectively (see Eq. \ref{eqn:geffs}).
Following \citet{KippenhahnThomas}, we define the equivalent radius of the equipotential ellipsoid, $r_P$, as:
\begin{equation}
\label{eqn:v_P}
v_P = \frac{4 \pi}{3} r_P^3,
\end{equation}
where $v_P$ is the volume enclosed by an isobar. 
For any quantity $q$ that is not uniform across an isobar, its mean value is defined as:
\begin{equation}
\label{eqn:q} 
\langle q \rangle = \frac{1}{S_P}\int_{\Psi = \mathrm{const}}^{} q d\sigma,
\end{equation}
where $S_P$ denotes the surface of the isobar
\begin{equation}
\label{eqn:sp} 
S_P = \int_{\Psi}^{} d\sigma = \int_{\Psi}^{} r^2 \sin \theta d\theta d\phi,
\end{equation}
$d\sigma$ is an element of the surface, and $\phi$ denotes the azimuthal angle.

In shellular rotation, the standard equations of stellar structure (i.e., conservation of mass, momentum, and energy, together with energy transport) can be recovered from the modified set of equations by assuming that all quantities depend not on the local values of $\rho$ and $T$, but on their volume-averaged means between two isobars, $\overline{\rho}$ and $\overline{T}$. 
The full expression for $\overline{\rho}$ can be found in \citet{MEYNET97}.
For simplicity, we omit the overline notation for $\overline{\rho}$ and $\overline{T}$ in the remainder of this appendix.

Eq.~\ref{eqn:momentumRot} can be extended to the more general case in which accelerations cannot be neglected and the star 
is not in hydrostatic equilibrium: 
\begin{equation}
\frac{\partial P}{\partial m_P} = -\frac{1}{4\pi r_P^2}\frac{\partial u_P}{\partial t} - \frac{Gm_p}{4\pi r_P^4}f_P.
\end{equation}

Note that only the momentum conservation and energy transport equations are affected by the correction factors $f_P$ and $f_T$.

\subsection{Calculation of scalar quantities $f_P$ and $f_T$}
\begin{figure}
\centering
\includegraphics[width=0.5\textwidth]{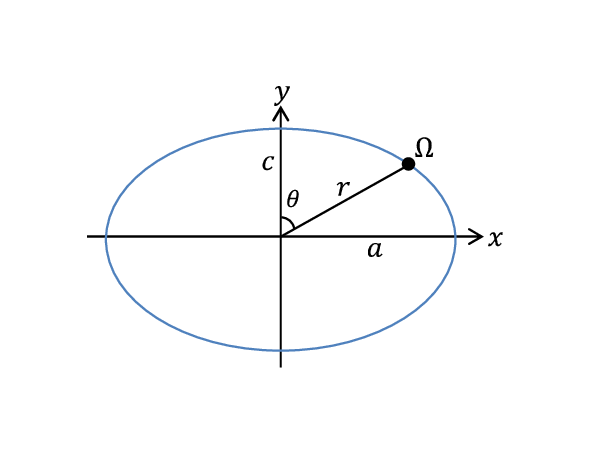}
\caption{Approximation of the equipotential geometry. The solid curve represents a symmetric rotational ellipsoid with a semi-major axis $a$ and a semi-minor axis $c$.}
\label{fig:Spheroid}
\end{figure}

The centrifugal forces generated in rotating stars lead to deviations from spherical symmetry. \citet{TASSOUL78} showed that for moderate or slow rotation rates (i.e., rotational velocities of $\sim 200$ km s$^{-1}$, corresponding to angular velocities of $\sim 10^{-5}$ rad s$^{-1}$) these deformations remain symmetric about the rotation axis (see Figure~\ref{fig:Spheroid}). Triaxial deformations occur only when the rotational energy becomes a significant fraction of the star's binding energy \citep{HEGER00}.
Under these assumptions, the star's isobaric surfaces can be approximated by symmetric rotational ellipsoids (see Figure~\ref{fig:Spheroid}). Consequently, the computation of the scalar quantities $f_P$ and $f_T$ requires determining the semi-major axis $a$, the semi-minor axis $c$, the radius $r(\theta, r_P)$, the surface area of the ellipsoid $S_P$, and the mean values $\langle g_{\mathrm{eff}} \rangle$ and $\langle g_{\mathrm{eff}}^{-1} \rangle$. 
The equivalent radius of the equipotential ellipsoid, $r_P$ (see Eq.~\ref{eqn:v_P}), can be obtained by solving the modified stellar structure equations 
for rotating stars (Eqs.~\ref{eqn:massRot}, \ref{eqn:momentumRot}, \ref{eqn:energyRot}, and \ref{eqn:luminosityRot}).
We begin by deriving an expression for $r(\theta, r_P)$. The radius and volume enclosed by a symmetric rotational ellipsoid are given by
\begin{equation}
v = \frac{4}{3}\pi a^2c \qquad ; \qquad r^2=\frac{a^2c^2}{a^2 \cos^2 \theta + c^2 \sin^2 \theta}.
\end{equation}
From the above expressions, and using the volume of the equivalent sphere (Eq.~\ref{eqn:v_P}), we obtain
\begin{equation}
\label{eqn:r_P^3}
r_P^3 = a^2c.
\end{equation}
The potential $\Psi_P$ is constant on an isobaric surface; therefore, the following condition must be satisfied:
\begin{equation}
\Psi_P(\theta=0) = \Psi_P(\theta=\pi/2).
\end{equation}
From Eq.~\ref{eqn:potential}, and noting that $r(\theta=0) = c$ and $r(\theta = \pi/2) = a$, we obtain
\begin{equation}
\label{eqn:Gm_Pc}
\frac{G m_P}{c} = \frac{G m_P}{a} + \frac{1}{2} \Omega^2 a^2.
\end{equation}
Solving the system formed by Eqs.~\ref{eqn:r_P^3} and \ref{eqn:Gm_Pc} yields the semi-major and semi-minor axes:
\begin{equation}
a = r_P \left( 1 - \frac{\Omega^2 r_P^3}{2 G m_P} \right)^{-1/3},
\qquad
c = r_P \left( 1 - \frac{\Omega^2 r_P^3}{2 G m_P} \right)^{2/3}.
\end{equation}

Once the radius $r$ is determined, the isobaric surface can be calculated from Eq.~\ref{eqn:sp}. Considering the symmetry with respect to the azimuthal angle $\phi$, this expression simplifies to:
\begin{equation}
\label{eqn:sp_Simp}
S_P = 4 \pi \int_{0}^{\pi/2} r^2 \sin \theta \, d\theta.
\end{equation}
There exist several formulas and approximations for computing the surface area of an ellipsoid (e.g., Knud Thomsen's formula). In this work, we calculate the surface area $S_P$ via numerical integration using the {\tt QUADPACK}\footnote{http://www.netlib.org/quadpack/}
 library, which provides very fast computation with an accuracy of approximately $10^{-8}$.

We now derive an expression for the modulus of the effective gravity, $g_{\mathrm{eff}}$, in order to compute $\langle g_{\mathrm{eff}} \rangle$ and $\langle g_{\mathrm{eff}}^{-1} \rangle$. In spherical coordinates, the first two components of the effective gravity, $g_{\mathrm{eff}} = (-g_{\mathrm{eff},r}, g_{\mathrm{eff},\theta}, 0)$, are given in the Roche model (see \citealt{MAEDER09}) by:
\begin{eqnarray}
\label{eqn:geffrtetha} 
g_{\mathrm{eff},r}&=&\frac{d\phi}{dr}-\Omega^2 r \sin^2\theta=\frac{Gm_P}{r^2}-\Omega^2 r \sin^2\theta \nonumber \\
g_{\mathrm{eff},\theta}&=&\Omega^2 r \sin\theta\cos\theta
\end{eqnarray} 
Consequently, the modulus of the effective gravity is
\begin{equation}
\label{eqn:geffmod}
g_{\mathrm{eff}} = \sqrt{\left( \frac{G m_P}{r^2} - \Omega^2 r \sin^2\theta \right)^2 + \left( \Omega^2 r \sin\theta \cos\theta \right)^2}.
\end{equation}
The mean values of the effective gravity over an isobaric surface, $\langle g_{\mathrm{eff}} \rangle$ and $\langle g_{\mathrm{eff}}^{-1} \rangle$, are defined as
\begin{eqnarray}
\label{eqn:geffs} 
\langle g_{\mathrm{eff}} \rangle &=& \frac{4\pi}{S_P}\int_{0}^{\frac{\pi}{2}} g_{\mathrm{eff}} r^2 \sin \theta d\theta \nonumber \\
\langle g_{\mathrm{eff}} \rangle^{-1} &=& \frac{4\pi}{S_P}\int_{0}^{\frac{\pi}{2}} \frac{1}{g_{\mathrm{eff}}} r^2 \sin \theta d\theta,
\end{eqnarray} 
and can be evaluated via numerical integration. For simplicity, we will denote $g_{\mathrm{eff}}$ simply as $g$ in the remainder of this text.

\subsection{Critical Angular Velocity}
\label{critical_angular_velocity}
In the Roche model, the critical angular velocity (also referred to as the break-up velocity) is reached when the modulus of the centrifugal force equals that of the gravitational attraction at the equator. Using Eq. \ref{eqn:geffmod}, we can determine the maximum angular velocity, $\Omega_{\mathrm{crit}}$, at which the effective gravity vanishes at the equator ($\theta = \pi/2$):
\begin{equation}
\label{eqn:wcrit} 
\Omega_{\mathrm{crit}}^2 =\frac{GM}{R_{\mathrm{e,crit}}^3},
\end{equation} 
Here, $R_{\mathrm{e,crit}}$ is the equatorial radius at break-up, and $M$ is the total stellar mass \citep{MAEDER09}. 
Substituting $\Omega_{\mathrm{crit}}$ into the surface equation at break-up (Eq. \ref{eqn:Gm_Pc}) and setting $a = R_{\mathrm{e,crit}}$ and $c = R_{\mathrm{p,crit}}$, we obtain the ratio of the equatorial to polar radius at the critical velocity:
\begin{equation}
\label{eqn:R_ec/R_pc} 
\frac{R_{\mathrm{e,crit}}}{R_{\mathrm{p,crit}}} = \frac{3}{2}.
\end{equation} 
Finally, the critical angular velocity for the 1.4 $M_{\odot}$ neutron star used in this work is:
\begin{equation}
\label{eqn:wcrit_value} 
\Omega_{\mathrm{crit}} \simeq 7550 \, \mathrm{rad} \cdot s^{-1} \quad (\mathrm{frequency} \simeq 1200 \, \mathrm{Hz} \, , \, \mathrm{period} \simeq 0.83 \, \mathrm{ms}).
\end{equation} 

It is important to stress that neutron stars are expected to rotate at some fraction of this critical angular velocity. Indeed, several mechanisms have been suggested to prevent a neutron star to achieve this critical value, including emission of gravitational waves, the presence of magnetic torques, and instabilities driven by viscosity, that can remove angular momentum before the star reaches break-up. In addition, the fastest rotating neutron stars ever observed do not exceed about 0.6-0.7 of the theoretical break-up limit (see, e.g., PSR J1748-2446ad, for which an angular velocity of 
4587 \, $\mathrm{rad} \cdot s^{-1}$ has been inferred; \citealt{Chakrabarty05}). Therefore, our choice of 0.8 $\Omega_{\mathrm{crit}}$ can be considered as a realistic upper limit for rotating neutron stars.

\subsection{Transport of Angular Momentum}
For shellular rotation, the equation of angular momentum transport in the vertical direction can be expressed, in Lagrangian formulation, as 
\citep{ZAHN92,MAEDER98}:
\begin{equation}
\centering
\label{eq:angular_momentum}
\rho\frac{\partial}{\partial t}\left(r^2\Omega\right)_{M_r}=\frac{1}{5r^2}\frac{\partial}{\partial r}\Big(\rho r^4\Omega U(r)\Big)+\frac{1}{r^2}\frac{\partial}{\partial r}\Big(\rho D r^4\frac{\partial \Omega}{\partial r}\Big).
\end{equation}
where $U(r)$ is the amplitude of the vertical component of the meridional circulation velocity, $u(r,\theta)$, at distance $r$.
 The coefficient $D$ represents the total vertical diffusion, which accounts for the transport of angular momentum by several instabilities, including convection, semi-convection, and shear turbulence.
Only the contribution of shear to the vertical turbulent diffusivity, $D_{\mathrm{s}}$, is considered here. According to \citet{ZAHN92}, shear instabilities are expected to dominate because they develop on a dynamical timescale, i.e., the shortest relevant timescale. The coefficient $D_{\mathrm{s}}$ is given by \citep{TALON97}:
\begin{equation}
\centering
D \simeq D_\mathrm{s}=\frac{8Ri_c}{5}\frac{\Big(r\frac{\partial\Omega}{\partial r}\Big)^2}{\frac{N_T^2}{K+D_\mathrm{h}}+\frac{N_{\mu}^2}{D_\mathrm{h}}}
\end{equation}
Here, $N_T^2=\frac{g\delta}{H_P}(\nabla_{\mathrm{ad}}-\nabla)$ and $N_{\mu}^2=\frac{g\varphi}{H_P}\nabla_{\mu}$; $K=\frac{4acT^3}{3k\rho^2C_P}=\frac{\chi}{\rho C_P}$ is the radiative diffusivity, with $\chi$ the radiative conductivity;  
$C_P = \left(\frac{\partial u}{\partial T}\right)_P + P\left(\frac{\partial v}{\partial T}\right)_P$ is the specific heat at constant pressure ($u$ and $v$ are the internal energy and specific volume, respectively);  
$H_P = -\frac{\mathrm{d}r}{\mathrm{d}\ln P} = -P \frac{\mathrm{d}r}{\mathrm{d}P}$ is the pressure scale height; 
and $Ri_c \approx 1/4$ is the critical Richardson number.  
Suitable definitions for the factors $\delta$ and $\varphi$ are given below (see Eq.~\ref{eq:delta_varphi}).
The adiabatic gradient, the actual temperature gradient, and the mean molecular weight gradient are defined, respectively, by \citep{HEGER00,kippenhahn1989stellar}:
\begin{equation}
\label{eq:nablas}
\nabla_{\mathrm{ad}} \equiv 
\left(\frac{\partial \ln T}{\partial \ln P}\right)_{\mathrm{ad}}
= \left(\frac{P}{T}\frac{\mathrm{d}T}{\mathrm{d}P}\right)_{\mathrm{ad}}
= -\frac{P}{C_P}\frac{\left(\frac{\partial P}{\partial T}\right)_v}{\left(\frac{\partial P}{\partial v}\right)_T},
\qquad
\nabla \equiv \frac{\mathrm{d}\ln T}{\mathrm{d}\ln P},
\qquad
\nabla_{\mu} \equiv \frac{\mathrm{d}\ln \mu}{\mathrm{d}\ln P}.
\end{equation}
In a radiative zone, the actual temperature gradient $\nabla$ equals the radiative gradient,
\[
\nabla_{\mathrm{rad}} = \frac{3}{16 \pi a c G} \, \frac{\kappa L P}{M T^4}.
\]
Factors $\delta$ and $\varphi$ are defined as follows \citep{HEGER00}:
\begin{eqnarray}
\label{eq:delta_varphi}
\delta=-\left(\frac{\partial\mathrm{ln}\rho}{\partial \mathrm{ln}T}\right)_{P,\mu} = \frac{T}{v}\left(\frac{\partial v}{\partial T}\right)_{P,\mu}=-T\rho\frac{\left (\frac{\partial P}{\partial T}\right)_{v,\mu}}{\left (\frac{\partial P}{\partial v}\right)_{T,\mu}} \nonumber \\
\varphi=\left(\frac{\partial \mathrm{ln}\rho}{\partial \mathrm{ln}\mu}\right)_{P,T}=-\frac{\mu}{v}\left(\frac{\partial v}{\partial \mu}\right)_{P,T} = \frac{\mu}{v} \frac{(\frac{\partial P}{\partial \mu})_{T,v}}{\left (\frac{\partial P}{\partial v}\right)_{T,\mu}}.
\end{eqnarray}
$D_\mathrm{h}$ is the viscosity coefficient associated with horizontal turbulence. We adopt the prescription of \citet{MAEDER03}:
\begin{equation}
\label{eq:Dh}
\centering
D_\mathrm{h} = \left(\frac{3}{400 n \pi}\right)^{1/3}
\, r \left[ r \, \Omega(r) \, V(r) \, \big(2V(r) - \alpha U(r)\big) \right]^{1/3},
\end{equation}
where $\alpha=\frac{1}{2}\frac{\mathrm{d} \, \mathrm{ln}(r^2\Omega)}{\mathrm{d} \, \mathrm{ln}r}$,
and $n$ is a parameter that can take the values $1$, $3$, or $5$ (with $n=1$ adopted in this work).

\subsection{Meridional Circulation}
Meridional circulation arises because equipotential surfaces are closer near the poles and farther apart near the equator as a result of the centrifugal force. Since the radiative flux is proportional to the effective gravity (i.e., to the spacing between equipotentials), there is an excess of flux along the polar axis and a deficit near the equatorial plane. This thermal imbalance drives large-scale circulation currents in the meridional plane.
The most widely adopted expression for the amplitude of the meridional circulation velocity, $U(r)$, is given by \citep{MAEDER98, MAEDER09}:

\begin{equation}
\centering
\label{eq:U}
U(r)=\frac{P}{\overline{\rho}\overline{g}C_P\overline{T}[\nabla_{\mathrm{ad}}-\nabla+(\varphi/\delta)\nabla_{\mu}]}\Big\{\frac{L}{M_{\star}}(E_{\Omega}+E_{\mu})+\frac{C_P T}{\delta}\frac{\partial \Theta}{\partial t}\Big\}
\end{equation}
Here, $M_{\star}=M(1-\frac{\Omega^2}{2\pi G\rho_m})$ is the reduced mass 
(i.e., the effective mass that would produce the same gravitational attraction in the absence of centrifugal force, where $\rho_m$ is the mean density inside the corresponding surface \citep{MAEDER03}. The quantity $\overline{g}$ denotes the mean \textit{effective} gravity under the assumption of shellular rotation. Similarly, $\overline{\rho}$ is the density averaged over the volume between two isobars. For simplicity, we omit the overline notation hereafter.
The terms $E_{\Omega}$ and $E_{\mu}$, which depend on the rotation profile and the mean molecular weight, are given by:
\begin{eqnarray}
\label{eq:Eomega}
E_{\Omega}&=&\frac{8\Omega^2 r^3}{3GM}\Big[1-\frac{\Omega^2}{2\pi G \rho}-\frac{\overline{\epsilon}+\overline{\epsilon}_{grav}}{\epsilon_m}\Big] \nonumber \\
&-&\frac{\rho_m}{\rho}\Big\{\frac{r}{3}\frac{\partial}{\partial r}\Big[H_T\frac{\partial}{\partial r}(\frac{\Theta}{\delta})-\chi_T(\frac{\Theta}{\delta})+(1-\frac{1}{\delta})\Theta\Big] \nonumber \\
&-&\frac{2H_T}{r}(1+\frac{D_\mathrm{h}}{K})(\frac{\Theta}{\delta})+\frac{2}{3}\Theta\Big\} \nonumber \\
&-&\frac{\overline{\epsilon}+\overline{\epsilon}_{grav}}{\epsilon_m}\Big[H_T\frac{\partial}{\partial r}(\frac{\Theta}{\delta})+(f_{\epsilon}\epsilon_T-\chi_T)(\frac{\Theta}{\delta}) \nonumber \\
&+&(2-f_{\epsilon}-\frac{1}{\delta})\Theta\Big]
\end{eqnarray}
and
\begin{eqnarray}
\label{eq:Emu}
E_{\mu}&=&\frac{\rho_m}{\rho}\Big\{\frac{r}{3}\frac{\partial}{\partial r}\Big[H_T\frac{\partial}{\partial r}(\frac{\varphi}{\delta}\Lambda)-(\chi_{\mu}+\frac{\varphi}{\delta}\chi_T+\frac{\varphi}{\delta})\Lambda\Big] \nonumber \\
&-&\frac{2H_T}{r}(\frac{\varphi}{\delta}\Lambda)\Big\} \nonumber \\
&+&\frac{\overline{\epsilon}+\overline{\epsilon}_{grav}}{\epsilon_m}\Big\{H_T\frac{\partial}{\partial r}(\frac{\varphi}{\delta}\Lambda)+ \Big[ f_{\epsilon}(\epsilon_{\mu}+\frac{\varphi}{\delta}\epsilon_T)-\chi_{\mu} \nonumber \\
&-&\frac{\varphi}{\delta}(\chi_T+1) \Big] \Lambda\Big\},
\end{eqnarray}
Here, $H_T = -\frac{\mathrm{d}r}{\mathrm{d}\ln T} = -T \frac{\mathrm{d}r}{\mathrm{d}T}$ 
is the temperature scale height, and 
\[
f_{\epsilon} \equiv \frac{\overline{\epsilon}}{\overline{\epsilon} + \overline{\epsilon}^{\mathrm{grav}}},
\]
with $\overline{\epsilon}$ and $\overline{\epsilon}^{\mathrm{grav}}$ denoting the mean nuclear energy generation rate per unit mass and the mean release rate of gravitational energy per unit mass, respectively. $\rho_m=\frac{3\mathrm{M}}{4\pi r^3}$ is the mean density inside the considered level surface and $\epsilon_m(r)\equiv L(r)/M(r)$ \citep{DENISSENKOV99}. 
The release rate of gravitational energy can be expressed, following \citet{kippenhahn1989stellar}, as
\begin{equation}
\overline{\epsilon}^{\mathrm{grav}} = - C_P \frac{\partial T}{\partial t} + \frac{\delta}{\rho} \frac{\partial P}{\partial t}.
\end{equation}
Finally, $\chi_{\mu}$ and $\epsilon_{\mu}$ denote the logarithmic derivatives of the radiative conductivity $\chi$ and the nuclear energy generation rate $\epsilon$ with respect to the mean molecular weight $\mu$, while derivatives with respect to $T$ are written as $\chi_T$ and $\epsilon_T$. These quantities are defined as:
\begin{equation}
\chi_T=\left(\frac{\partial \mathrm{ln} \chi}{\partial \mathrm{ln} T}\right)_{\mu,P}=3-\frac{T}{k}\frac{\partial k}{\partial T}+\delta \qquad \chi_{\mu}=\left(\frac{\partial \mathrm{ln} \chi}{\partial \mathrm{ln} \mu}\right)_{T,P} =
-\frac{\mu}{k}\frac{\partial k}{\partial \mu}-\varphi \nonumber \\
\end{equation}
\begin{equation}
\epsilon_T=\left(\frac{\partial \mathrm{ln} \epsilon}{\partial \mathrm{ln} T}\right)_{\mu,P} \qquad 
\epsilon_{\mu}=\left(\frac{\partial \mathrm{ln} \epsilon}{\partial \mathrm{ln} \mu}\right)_{T,P}.
\end{equation}

The quantities $\Theta$ and $\Lambda$ represent, respectively, the relative variations of density and mean molecular weight over an isobar, and are defined as
\begin{equation}
\label{eq:Theta}
\Theta = \frac{\tilde{\rho}}{\overline{\rho}} = \frac{1}{3} \frac{r^2}{\overline{g}} \frac{\partial \Omega^2}{\partial r}, 
\qquad 
\Lambda = \frac{\tilde{\mu}}{\overline{\mu}}.
\end{equation}
Here, $\overline{\rho}$ and $\overline{\mu}$ denote the mean values on the isobar, while $\tilde{\rho}$ and $\tilde{\mu}$ indicate the horizontal variations.  

The time derivative of $\Lambda$ is given by
\begin{equation}
\label{eq:Lambda_time_derivative}
\frac{\partial \Lambda}{\partial t} = \frac{U}{H_P} \nabla_{\mu} - \frac{6}{r^2} D_\mathrm{h} \Lambda.
\end{equation}

The continuity equation provides a relation between the amplitudes of the horizontal, $V(r)$, and vertical, $U(r)$, components of the meridional circulation, as derived by \citet{ZAHN92}:
\begin{equation}
\centering
\label{eq:V_dr}
\frac{1}{r}\frac{\partial}{\partial r}[\rho r^2 U(r)]-6\rho V(r)=0.
\end{equation}

\subsection{Mixing and Transport of Chemical Elements}
Shear in differentially rotating stars generates instabilities that drive the transport of chemical elements. The combined effect of meridional circulation and horizontal turbulence on the transport of nuclides can be approximated as a diffusion process. Following \citet{TALON97}, the diffusion of chemical elements can be treated as:
\begin{equation}
\centering
\label{eq:diffusion_of_chemicals}
\rho \frac{\partial c_i}{\partial t} = \dot{c}_i|_{\mathrm{nuc}} + \frac{1}{r^2} \frac{\partial}{\partial r} \left[ r^2 \rho \left( D_\mathrm{eff} + D_\mathrm{s} \right) \frac{\partial c_i}{\partial r} \right],
\end{equation}
where $c_i$ is the concentration of species $i$, and $\dot{c}_i|_{\mathrm{nuc}}$ represents the nuclear production or destruction rate, as provided by the stellar structure code.  Alternatively, the formulation can be expressed as (see \citealt{MEYNET00}):
\begin{equation}
\centering
\label{eq:Xidt}
\left. \frac{\partial X_i}{\partial t}\right\vert_{m_r}=\left. \frac{\partial X_i}{\partial t}\right\vert_{nuc}+\left. \frac{\partial X_i}{\partial t}\right\vert_{cir}=\left. \frac{\partial X_i}{\partial t}\right\vert_{nuc}+\frac{\partial}{\partial m}\left[(4\pi r^2 \rho)^2(D_\mathrm{eff}+D_\mathrm{s})\frac{\partial X_i}{\partial m}\right]
\end{equation}
Here, $\left. \frac{\partial X_i}{\partial t} \right|_{\mathrm{cir}}$ represents the change in the mass fraction $X_i$ due solely to meridional circulation. Equation \ref{eq:Xidt} is subject to the following boundary conditions \citep{MEYNET04}: 
\begin{eqnarray}
\label{eq:Xidt_bc}
\left. \frac{\partial X_i}{\partial m} \right|_{m_r=0} = 
\left. \frac{\partial X_i}{\partial m} \right|_{m_r=M_0} = 0.
\end{eqnarray}
where $M_0$ denotes the total stellar mass.

The effective diffusivity coefficient, $D_\mathrm{eff}$, following \citet{CHABOYER92}, is given by
\begin{equation}
\centering
\label{eq:Deff}
D_\mathrm{eff} = \frac{|r U(r)|^2}{30 D_\mathrm{h}}.
\end{equation}

\subsection{The Full Set of Equations}
In summary, shellular rotation has been implemented in this work by the following set of equations: 
\begin{equation}
\Theta=\frac{1}{3}\frac{r^2}{g}\frac{\partial \Omega^2}{\partial r}
\end{equation} 

\begin{eqnarray}
\label{eq:U_dr}
U&=&\frac{P}{\rho g C_P\overline{T}[\nabla_{\mathrm{ad}}-\nabla+(\varphi/\delta)\nabla_{\mu}]}\bigg\{\frac{L}{M_{\star}}\Big[\frac{8\Omega^2 r^3}{3\mathrm{GM}}\Big(1-\frac{\Omega^2}{2\pi G \rho}-\frac{\overline{\epsilon} + \overline{\epsilon}_{grav}}{\epsilon_m}\Big)\nonumber \\
&-&\frac{\rho_m}{\rho}\left(\frac{r}{3}\frac{\partial}{\partial r}\mathrm{A}+\left[-\frac{2H_T}{r}\left(1+\frac{D_\mathrm{h}}{K}\right)+\frac{2\delta}{3}\right]\left(\frac{\Theta}{\delta}\right)+\frac{2H_T}{r}\left(\frac{\varphi}{\delta}\Lambda\right)\right) \nonumber \\
&-&\frac{\overline{\epsilon}+\overline{\epsilon}_{grav}}{\epsilon_m}\left(\mathrm{A}+f_{\epsilon}\epsilon_T\left(\frac{\Theta}{\delta}\right)+(1-f_{\epsilon})\Theta-\left(f_{\epsilon}\epsilon_{\mu}+f_{\epsilon}\frac{\varphi}{\delta}\epsilon_T\right)\Lambda\right) \nonumber \\
&-&\frac{\Omega^2}{2\pi G\rho}\Theta\Big]+\frac{C_P T}{\delta}\frac{\partial \Theta}{\partial t}\bigg\}
\end{eqnarray}
\begin{equation}
\label{eq:angular_momentum_dr}
\rho\frac{\partial}{\partial t}\left(r^2\Omega\right)=\frac{1}{5r^2}\frac{\partial}{\partial r}\Big(\rho r^4\Omega U\Big)+\frac{1}{r^2}\frac{\partial}{\partial r}\Big(\rho D_\mathrm{s} r^2g\frac{3}{2}\frac{\Theta}{\Omega}\Big)
\end{equation}
with
$\mathrm{A}=H_T\frac{\partial}{\partial r}\left(\frac{\Theta}{\delta}-\frac{\varphi}{\delta}\Lambda\right)-(\chi_T+1-\delta)\frac{\Theta}{\delta}+\left(\chi_{\mu}+\frac{\varphi}{\delta}\chi_T+\frac{\varphi}{\delta}\right)\Lambda$

\begin{equation}
\frac{\partial \Lambda}{\partial t}=\frac{U}{H_P}\nabla_{\mu}-\frac{6}{r^2}D_\mathrm{h}\Lambda,
\end{equation}
where $\Theta=\frac{2}{3}\frac{r^2}{\overline{g}} \Omega \frac{\partial \Omega}{\partial r}$ is used 
to avoid introducing a second-order derivative in Eq.~\ref{eq:angular_momentum_dr}. The Gratton-\"Opik term, 
$-\frac{\Omega^2}{2\pi G\rho}\Theta$ 
\citep{Gratton45, OPIK51},
has been added to Eq.~\ref{eq:U_dr} following \citet{DENISSENKOV99}. Near the stellar surface,
the factor $\frac{\Omega^2}{2\pi G\rho}$ 
becomes significant and can even reverse the direction of circulation, as first noted by \citet{Gratton45} and \citet{OPIK51}. This reversal creates an outer circulation cell that transports angular momentum outward. Known as the Gratton-\"Opik circulation cell, it plays a key role in stellar evolution by transferring angular momentum inward and enhancing surface rotation \citep{MAEDER03}.

\subsection{Boundary Conditions}
\label{CH3_Boundary_conditions}
The system of equations requires a corresponding set of boundary conditions. At the stellar surface, we adopt the following conditions \citep{TALON97, DENISSENKOV99}:
\begin{equation}
\frac{\partial \Omega}{\partial m} = 0 \qquad \mathrm{and} \qquad U(m)=0
\end{equation}
while at the edge of the core we assume
\begin{equation}
\label{boundary_conditions_core}
\frac{\partial \Omega}{\partial m} = 0 \qquad \mathrm{and} \qquad \Omega(m) = \Omega_\mathrm{c}
\end{equation}
where $\Omega_c$ denotes the rotation velocity at the core. The following conditions follow directly from the above:
\begin{equation}
\Theta = 0 \qquad \mathrm{and} \qquad \alpha = 1
\end{equation}
at the stellar surface and at the core.

We now calculate the temporal evolution of $\Omega_c$. 
Applying condition~\ref{boundary_conditions_core} to Eq.~\ref{eq:angular_momentum} yields:
\begin{equation}
\centering
\label{eq:angular_momentumBC1}
\rho 5 r^2 \frac{\partial}{\partial t}\left(r^2\Omega_\mathrm{c}\right)=\frac{\partial}{\partial r}\Big(\rho r^4\Omega U(r)\Big).
\end{equation}
Since the relaxation time to reach a state of stationary rotation is much shorter than the stellar lifetime, we conclude that
$\frac{\partial \Omega}{\partial t} \gg \frac{\partial r}{\partial t}$, and then
\begin{equation}
\centering
\label{eq:omega_core1}
\frac{\partial \Omega_\mathrm{c}}{\partial t}=\frac{\left( \rho r^4\Omega U(r) \right)_{r=r_\mathrm{c}}}{5 \int_{0}^{r_\mathrm{c}} \rho r^4 dr}
\end{equation}
where $r_\mathrm{c}$ is the core radius. This expression for $\Omega_\mathrm{c}$ is identical to that obtained by \citet{DENISSENKOV99}, under the assumptions that the core rotates as a rigid body and that the star conserves its angular momentum (i.e., mass loss is neglected).
With these assumptions, we calculate $\int_{0}^{r_\mathrm{c}} \rho r^4 dr$. The core mass $M_\mathrm{c}$ can be expressed in terms of the core radius, $r_\mathrm{c}$, and the core density,  $\rho_\mathrm{c}$:
\begin{equation}
\centering
\label{eq:mc}
M_\mathrm{c}=\frac{4 \pi r_\mathrm{c}^3}{3}\rho_\mathrm{c} \qquad \mathrm{and} \qquad r_\mathrm{c}=\left(\frac{3M_\mathrm{c}}{4 \pi \rho_\mathrm{c}}\right)^{\frac{1}{3}}.
\end{equation}
Taking into account that $dr=\frac{dm}{4 \pi r^2 \rho}$, the integral can be solved by applying a change of variables:
\begin{equation}
\centering
\label{eq:integral}
\int_{0}^{r_\mathrm{c}} \rho r^4 dr=\frac{1}{4 \pi} \int_{0}^{\left(\frac{3M_\mathrm{c}}{4 \pi \rho_\mathrm{c}}\right)^{\frac{1}{3}}} r^2dm=\frac{1}{4 \pi}\left[ \frac{r^3}{3}\frac{dm}{dr}\right]_0^{\left(\frac{3M_\mathrm{c}}{4 \pi \rho_\mathrm{c}}\right)^{\frac{1}{3}}}=\frac{r^2M_\mathrm{c}}{4 \pi},
\end{equation}
and finally,
\begin{equation}
\centering
\label{eq:omega_core2}
\frac{\partial \Omega_\mathrm{c}}{\partial t}=\frac{4 \pi}{5}\frac{\left( \rho r^2\Omega U(r) \right)_{r=r_\mathrm{c}}}{M_\mathrm{c}}.
\end{equation}

The complete set of equations for shellular rotation, together with the corresponding boundary conditions, can be solved using a relaxation technique, such as the {\it Henyey method} \citep{Henyey64}.

\section{Main Properties of the Models Computed}

Table \ref{tab1} summarizes the main properties of the five models computed in this work, along five bursting sequences. $\rho_{\mathrm{max,ign}}$ and $P(t_{\mathrm{max}})$ correspond to the maximum density and pressure attained at the ignition shell (defined as the first shell that reaches $T > 5 \times 10^{8}$ K, for each burst) at time $t_{\mathrm{max}} \ (s)$. $T(t_{\mathrm{max}})$ is the temperature at the envelope base at that instant. For the first burst, the ignition shell corresponds to the innermost envelope shell, therefore $T(t_{\mathrm{max}})$ is the ignition temperature at the envelope base for the first burst,
while $\rho_{\mathrm{max,ign}}=\rho(t_{\mathrm{max}})$ and $P(t_{\mathrm{max}})=P_{\mathrm{max,ign}}$. $P_{\mathrm{stabil}}$ is the stabilized pressure, i.e. the value of the pressure when $T_{\mathrm{\mathrm{base}}} = 6 \ \times \ 10^8$ K. $\Delta z(t_{\mathrm{max}})$ is the size of the envelope at $t_{\mathrm{max}}$, and $\Delta z_{\mathrm{peak}}$ is the maximum expansion of the envelope achieved during the burst. $\tau_{\mathrm{rec}}$ is the recurrence time (i.e., time between two consecutive bursts, calculated from peak luminosities, L$_{\mathrm{peak}}$/L$_\odot$), and $\alpha$ is the ratio between persistent and burst luminosities. We define $\alpha = \int_{t+\tau_{\mathrm{rec}}}^{t}L(t)dt / \int_{t'+\tau_{\mathrm{0.01}}}^{t'}L(t)dt$, with the latter term integrated over the time during which the burst exceeds 1\% of its peak
luminosity, $\tau_{\mathrm{0.01}}$. Note that during the interburst period, the accretion luminosity, $L_{\mathrm{acc}} = GM\dot{M}/R \sim 1.5 \times 10^{37}$ erg s$^{-1}$, will hide the thermal emission from the cooling ashes.

\startlongtable
\begin{deluxetable*}{lccccc}
\tablecaption{Properties of all models computed in this work, along five bursting sequences}\label{tab1}
\startdata
	&	& & Model 1		&	&	\\[0.5ex]
\hline \\[-1.8ex]
Burst sequence  & 1 & 2 & 3 & 4 & 5 \\
$\rho_{\mathrm{max,ign}}$ (g cm$^{-3}$)	&	3.4 $\times$ 10$^5$	&	7.1 $\times$ 10$^5$	&	5.9 $\times$ 10$^5$	&	6.3 $\times$ 10$^5$	&	5.9 $\times$ 10$^5$\\[0.5ex]
$P(t_{\mathrm{max}})$ (dyn cm$^{-2}$)	&	1.2 $\times$ 10$^{22}$	&	1.7 $\times$ 10$^{22}$ 	&	1.4 $\times$ 10$^{22}$	&	1.4 $\times$ 10$^{22}$	&	1.3 $\times$ 10$^{22}$\\[0.5ex]
$P_{\mathrm{stabil}}$ (dyn cm$^{-2}$)	&	1.19 $\times$ 10$^{22}$	&	2.07 $\times$ 10$^{22}$ 	&	3.13 $\times$ 10$^{22}$	&	4.18 $\times$ 10$^{22}$	&	5.22 $\times$ 10$^{22}$\\[0.5ex]
$T(t_{\mathrm{max}})$ (K)	&	2.7 $\times$ 10$^{8}$	&	2.4 $\times$ 10$^8$	&	2.3 $\times$ 10$^8$	&	2.2 $\times$ 10$^8$	&	2.1 $\times$ 10$^8$\\[0.5ex]
$\Delta z (t_{\mathrm{max}})$ (m)	&	12.8			&	 12.6	&	12.7	&	12.8	&	12.7\\[0.5ex]
t$_{\mathrm{max}}$ (s)	&	21006	&	36146 	&	53871	&	71346	&	87279\\[0.5ex]
$\tau_{\mathrm{rec}}$ (hr) &	-	&	4.4 	&	5.2	&	5.1	&	5.1\\[0.5ex]
$\alpha$ &	-	&	26	&	31	&	31	&	31\\[0.5ex]
$\Delta z_{\mathrm{peak}}$ (m)	&	44.8	&	 44.7	&	43.7		&	39.9	&	38.0\\[0.5ex]
L$_{\mathrm{peak}}$/L$_\odot$	&	1.0 $\times$ 10$^{5}$	&	 2.1 $\times$ 10$^{5}$	&	2.0 $\times$ 10$^{5}$	&	1.7 $\times$ 10$^{5}$	&	1.6 $\times$ 10$^{5}$\\[0.5ex]
\hline \\[-1.8ex]
	&	& & Model 2		&	&	\\[0.5ex]
\hline \\[-1.8ex]
Burst sequence  & 1 & 2 & 3 & 4 & 5 \\
$\rho_{\mathrm{max,ign}}$ (g cm$^{-3}$)	&	3.3 $\times$ 10$^5$	&	6.6 $\times$ 10$^5$	&	6.0 $\times$ 10$^5$	&	6.1 $\times$ 10$^5$	&	5.8 $\times$ 10$^5$\\[0.5ex]
$P(t_{\mathrm{max}})$ (dyn cm$^{-2}$)	&	1.2 $\times$ 10$^{22}$	&	1.5 $\times$ 10$^{22}$	&	1.4 $\times$ 10$^{22}$	&	1.4 $\times$ 10$^{22}$	&	1.3 $\times$ 10$^{22}$\\[0.5ex]
$P_{\mathrm{stabil}}$ (dyn cm$^{-2}$)	&	1.17 $\times$ 10$^{22}$	&	2.03 $\times$ 10$^{22}$ 	&	3.05 $\times$ 10$^{22}$	&	4.04 $\times$ 10$^{22}$ 	&	5.04 $\times$ 10$^{22}$\\[0.5ex]
$T(t_{\mathrm{max}})$ (K)	&	2.7 $\times$ 10$^{8}$	&	2.4 $\times$ 10$^8$	&	2.3 $\times$ 10$^8$	&	2.2 $\times$ 10$^8$	&	2.1 $\times$ 10$^8$\\[0.5ex]
$\Delta z (t_{\mathrm{max}})$ (m)	&	12.8 		&	12.7 	&	12.8	&	12.8	&	12.8\\[0.5ex]
t$_{\mathrm{max}}$ (s)	&	20869	&	 35830	&	53190	&	69554	&	85763\\[0.5ex]
$\tau_{\mathrm{rec}}$ (hr) &	-	&	 4.3 	&	5.1	&	4.9	&	5.0\\[0.5ex]
$\alpha$ &	-	&	27 	&	32	&	31	&	32\\[0.5ex]
$\Delta z_{\mathrm{peak}}$	&	45.9	&	 43.4 	&	43.7	&	41.1	&	40.3\\[0.5ex]
L$_{\mathrm{peak}}$/L$_\odot$ (m)	&	1.0 $\times$ 10$^{5}$	&	 2.0 $\times$ 10$^{5}$	&	1.6 $\times$ 10$^{5}$	&	1.5 $\times$ 10$^{5}$	&	1.8 $\times$ 10$^{5}$\\[0.5ex]
\hline \\[-1.8ex]
	&	& & Model 3		&	&	\\[0.5ex]
\hline \\[-1.8ex]
Burst sequence  & 1 & 2 & 3 & 4 & 5 \\
$\rho_{\mathrm{max,ign}}$ (g cm$^{-3}$)	&	2.9 $\times$ 10$^5$	&	6.0 $\times$ 10$^5$	&	5.8 $\times$ 10$^5$	&	5.5 $\times$ 10$^5$	&	5.5 $\times$ 10$^5$\\[0.5ex]
$P(t_{\mathrm{max}})$ (dyn cm$^{-2}$)	&	1.1 $\times$ 10$^{22}$	&	1.4 $\times$ 10$^{22}$	&	1.3 $\times$ 10$^{22}$	&	1.2 $\times$ 10$^{22}$	&	1.2 $\times$ 10$^{22}$\\[0.5ex]
$P_{\mathrm{stabil}}$ (dyn cm$^{-2}$)	&	1.09 $\times$ 10$^{22}$	&	1.88 $\times$ 10$^{22}$	&	2.81 $\times$ 10$^{22}$	&	3.73 $\times$ 10$^{22}$	&	4.69 $\times$ 10$^{22}$\\[0.5ex]
$T(t_{\mathrm{max}})$ (K)	&	2.6 $\times$ 10$^{8}$	&	2.5 $\times$ 10$^8$	&	2.3 $\times$ 10$^8$	&	2.2 $\times$ 10$^8$	&	2.1 $\times$ 10$^8$\\[0.5ex]
$\Delta z (t_{\mathrm{max}})$ (m)	&	13.3		&	13.4 	&	13.5	&	13.4	&	13.5\\[0.5ex]
t$_{\mathrm{max}}$ (s)	&	20508	&	35402 	&	52087	&	68154	&	85119\\[0.5ex]
$\tau_{\mathrm{rec}}$ (hr) &	-	&	 4.2	&	4.9	&	4.9	&	5.1\\[0.5ex]
$\alpha$ &	-	&	26 	&	31	&	31	&	32\\[0.5ex]
$\Delta z_{\mathrm{peak}}$ (m)	&	49.2	&	 48.5	&	46.1	&	45.9	&	42.6	\\[0.5ex]
L$_{\mathrm{peak}}$/L$_\odot$ &	9.4 $\times$ 10$^{4}$	&	 1.7 $\times$ 10$^{5}$	&	1.5 $\times$ 10$^{5}$	&	1.8 $\times$ 10$^{5}$	&	1.4 $\times$ 10$^{5}$\\[0.5ex]
\hline \\[-1.8ex]
	& &	& Model 4		&	&	\\[0.5ex]
\hline \\[-1.8ex]
Burst sequence  & 1 & 2 & 3 & 4 & 5 \\
$\rho_{\mathrm{max,ign}}$ (g cm$^{-3}$)	&	2.5 $\times$ 10$^5$	&	5.6 $\times$ 10$^5$	&	5.1 $\times$ 10$^5$	&	5.1 $\times$ 10$^5$	&	4.7 $\times$ 10$^5$\\[0.5ex]
$P(t_{\mathrm{max}})$ (dyn cm$^{-2}$)	&	9.2 $\times$ 10$^{21}$	&	1.3 $\times$ 10$^{22}$ 	&	1.1 $\times$ 10$^{22}$	&	1.1 $\times$ 10$^{22}$	&	9.5 $\times$ 10$^{21}$\\[0.5ex]
$P_{\mathrm{stabil}}$ (dyn cm$^{-2}$)	&	9.49 $\times$ 10$^{21}$	&	1.61 $\times$ 10$^{22}$ 	&	2.44 $\times$ 10$^{22}$	&	3.30 $\times$ 10$^{22}$	&	4.12 $\times$ 10$^{22}$\\[0.5ex]
$T(t_{\mathrm{max}})$ (K)	&	2.6 $\times$ 10$^{8}$	&	2.4 $\times$ 10$^8$	&	2.4 $\times$ 10$^8$	&	2.2 $\times$ 10$^8$	&	2.1 $\times$ 10$^8$\\[0.5ex]
$\Delta z (t_{\mathrm{max}})$ (m)	&	14.9		&	14.9	 	&	15.1		&	15.1	&	14.6\\[0.5ex]
t$_{\mathrm{max}}$ (s)	&	20130	&	 34266	&	51504	&	68703	&	83604\\[0.5ex]
$\tau_{\mathrm{rec}}$ (hr) &	-	&	4.0 	&	5.3	&	5.3	&	4.9\\[0.5ex]
$\alpha$ &	-	&	25 	&	31	&	33	&	30\\[0.5ex]
$\Delta z_{\mathrm{peak}}$ (m)	&	56.2	&	58.2 	&	55.7	&	53.3	&	49.7\\[0.5ex]
L$_{\mathrm{peak}}$/L$_\odot$ &	7.9 $\times$ 10$^{4}$	&	1.8 $\times$ 10$^{5}$ 	&	2.1 $\times$ 10$^{5}$	&	1.5 $\times$ 10$^{5}$	&	1.4 $\times$ 10$^{5}$\\[0.5ex]
\hline \\[-1.8ex]
	&	& & Model 5		&	&	\\[0.5ex]
\hline \\[-1.8ex]
Burst sequence  & 1 & 2 & 3 & 4 & 5 \\
$\rho_{\mathrm{max,ign}}$ (g cm$^{-3}$)	&	1.9 $\times$ 10$^5$	&	4.0 $\times$ 10$^5$	&	3.8 $\times$ 10$^5$	&	3.7 $\times$ 10$^5$	&	3.5 $\times$ 10$^5$\\[0.5ex]
$P(t_{\mathrm{max}})$ (dyn cm$^{-2}$)	&	6.8 $\times$ 10$^{21}$	&	8.4 $\times$ 10$^{21}$ 	&	7.2 $\times$ 10$^{21}$	&	6.9 $\times$ 10$^{21}$	&	6.1 $\times$ 10$^{21}$\\[0.5ex]
$P_{\mathrm{stabil}}$ (dyn cm$^{-2}$)	&	7.05 $\times$ 10$^{21}$	&	1.17 $\times$ 10$^{22}$ 	&	1.72 $\times$ 10$^{22}$	&	2.26 $\times$ 10$^{22}$	&	2.82 $\times$ 10$^{22}$\\\\[0.5ex]
$T(t_{\mathrm{max}})$ (K)	&	2.6 $\times$ 10$^{8}$	&	2.5 $\times$ 10$^8$	&	2.1 $\times$ 10$^8$	&	2.1 $\times$ 10$^8$	&	2.0 $\times$ 10$^8$\\[0.5ex]
$\Delta z (t_{\mathrm{max}})$ (m)	&	19.7		&	19.0 	&	17.5	&	17.5	&	16.7\\[0.5ex]
t$_{\mathrm{max}}$ (s)	&	19999	&	33074 	&	44962	&	59245	&	71929\\[0.5ex]
$\tau_{\mathrm{rec}}$ (hr) &	-	&	 3.8	&	4.4	&	4.4	&	4.4\\[0.5ex]
$\alpha$ &	-	&	26 	&	31	&	32	&	33\\[0.5ex]
$\Delta z_{\mathrm{peak}}$ (m)	&	74.4	&	76.6 	&	70.9	&	66.3	&	68.1\\[0.5ex]
L$_{\mathrm{peak}}$/L$_\odot$ &	5.4 $\times$ 10$^{4}$	&	 9.8 $\times$ 10$^{4}$	&	8.6 $\times$ 10$^{4}$	&	7.7 $\times$ 10$^{4}$	&	1.1 $\times$ 10$^{5}$\\[0.5ex]
\enddata
\end{deluxetable*}


\bibliography{paperv3}{}
\bibliographystyle{aasjournal}



\end{document}